\documentclass[12pt]{article}
\usepackage[centertags]{amsmath}
\usepackage{amsfonts}
\usepackage{amssymb}
\usepackage{amsthm}
\newcommand{\beq}{\begin{equation}}
\newcommand{\eeq}[1]{\label{#1}\end{equation}}
\newcommand{\bea}{\begin{eqnarray}}
\newcommand{\eea}[1]{\label{#1}\end{eqnarray}}

\begin{document}
\setlength{\topmargin}{-2cm} \setlength{\oddsidemargin}{0cm}
\setlength{\evensidemargin}{0cm}
\begin{titlepage}
\hfill CERN-TH/11-194\\
\vspace{20pt}
\begin{center}
{\Large \bf 4D Localization in Randall-Sundrum 2 Supergravity and in Vasiliev Theories}

\vspace{20pt}

{\large S. Ferrara$^{abc}$ and M. Porrati$^{d}$}

\vspace{12pt}

$^a$ Physics Department, Theory Unit, CERN\\
CH-1211, Geneva 23, Switzeland

\vspace{12pt}

$^b$ INFN Laboratori Nazionali di Frascati\\
Via Enrico Fermi 40, I 00044 Frascati, Italy

\vspace{12pt}

$^c$ Department of Physics and Astronomy, UCLA\\
Los Angeles, CA 90095-1547, USA

\vspace{12pt}

$^d$ Center for Cosmology and Particle Physics\\
Department of Physics\\ New York University\\
4 Washington Place\\ New York, NY 10003, USA

\end{center}
\vspace{20pt}

\begin{abstract}
We discuss the problem of localization of 4D massless states in Randall-Sundrum 2 (one-brane) models. A Randall-Sundrum 2 construction starting from N=8 gauged supergravity in 5D Anti de Sitter space gives rise to an N=4 supergravity-matter system. We explicitly show that only the modes of the N=4 graviton supermultiplet localize on the 4D brane, streamlining and generalizing previous works. We also point out that while charged 1/4 BPS black holes do exist in the 4D theory, they are always produced in sets of total charge zero. This zero-charge configuration uplifts to a 5D metric without naked singularities, thus avoiding the curvature singularity of the 5D uplift of an isolated charged BPS black hole. Finally, we resolve a puzzle with localization of massless high spin fields on a (putative) Randall-Sundrum  2 construction based on Vasiliev's high spin theories. We show that while high spin fields do localize, the gauge symmetry that ensures decoupling of their unphysical polarizations is anomalous. This implies that the high spin fields must acquire a mass.
\end{abstract}

\end{titlepage}

\newpage
\section{Introduction}
The one-brane Randall-Sundrum model (RS2)~\cite{rs2} is a compelling example of holographic duality with dynamical 
gravity~\cite{bnw,dl0,ahpr}. The 5D description of the model is semiclassical gravity plus matter in $AdS_5$ space cut off near the conformal boundary by a brane. The  brane tension is tuned so as to give a flat induced metric on its 4D world-volume. The key to having a dynamical graviton as opposed to a fixed external metric is that the 5D graviton admits a normalizable zero mode at zero mass, that is when the 4D momentum $k^\mu$ obeys $k^2=0$. 

When we write $AdS_5$  in Poincar\'e coordinates as
\beq
ds^2= {L^2 \over z^2} (dz^2 + dx^\mu dx_\nu), \qquad \mu,\nu=0,1,2,3,
\eeq{m1}
the RS2 brane is located at $z=\epsilon\ll L$ and $L$ is the AdS radius. Because of the symmetries of $AdS_5$, the radial wave function of the massless graviton is a product of a polarization tensor times a function of the radial coordinate $z$ and $k^2$. At $k^2=0$, this function becomes a pure power in $z$.  As it is well known, near the boundary $z=0$, the wave function of a field in $AdS_5$ behaves as a linear combination of two powers of $z$~\cite{agmo}. In the case of the graviton we have
\beq
g_{\mu\nu} ={L^2\over z^2} [\eta_{\mu\nu} + h_{\mu\nu}(x) + z^2 k_{\mu\nu}(x)].
\eeq{i2}
The term proportional to $z^{-2}$ is non-normalizable in $AdS_5$, but it becomes normalizable in RS2, where the space
is cutoff at $z=\epsilon$ and so it manifests as a dynamical 4D graviton. 

An alternative way to see that the graviton is dynamical is to compute the graviton Euclidean Green's function 
$G^{\mu\nu, \rho\sigma}(z,x| z',y)$ at $z=z'=\epsilon$. The Green's function contains contact terms diverging as inverse powers
of $\epsilon$; specifically, there is a term proportional to $L^3\epsilon^{-2}L_{\mu\nu,\rho\sigma}(x)\delta^4(x-y)$ (where $L_{\mu\nu,\rho\sigma}(x)=D_{x\,\mu} D_x^\mu +...$ is the kinetic term of a massless spin 2 field), which induces a kinetic
term for the boundary metric~\cite{bnw,dl0,ahpr}.
The 4D interpretation of this result is that the graviton kinetic term is induced by the self-energy arising from loops of a conformal field theory (CFT). 

When the bulk theory in 5D is a gauged supergravity with 32 supercharges (N=8) more fields propagate in the bulk and more localize on the 4D boundary. The holographic dual in this case is an N=4 superconformal field theory coupled
to ungauged N=4 supergravity. This setup requires certain bulk fields to localize, namely the graviton and its N=4 superpartners. Equally importantly, it requires certain fields {\em not} to localize! Most prominently, the $SU(4)$ gauged fields of 5D supergravity. For if they localized, they would generate dynamical gauge fields of an anomalous $SU(4)$ R symmetry that is incompatible with the flat space supersymmetry algebra. Ungauged N=4 4D supergravity does possess six graviphotons, but they originate from 5D antisymmetric forms. Because the gauge fields originate from two-forms and not vectors in 5D, they avoid the well known fact that vectors do not localize in RS2~\cite{bg,g} (see~\cite{ahpr} for
the 4D dual interpretation of this property). 

The mechanism responsible for localization of two-forms was discussed in~\cite{dl} in the context of N=2 supergravity; we shall review it in the next section, where it is shown that N=8 RS2 localizes all and only the fields needed to describe
an N=4 ungauged supergravity coupled to a superconformal field theory. Next section is also devoted to solving a puzzle raised in~\cite{fp}. There, it was noticed that because the 5D fields which originate the graviphotons of 4D N=4 supergravity descend from antisymmetric tensors in 5D, any 4D object charged under the graviphoton $U(1)^6$ is actually a string. Indeed, the explicit solution obtained by ``oxidation" of an extremal Reissner-Nordst\"om (RN) black hole, found in~\cite{clp}, describes a string extending all the way to $z=\infty$. The puzzle is that this solution has a naked singularity at $z=\infty$ (the Poincar\'e horizon of $AdS_5$). Here we propose a simple solution to this problem; namely: extremal (1/4 BPS) RN black holes are always produced by the N=4 4D supergravity plus superconformal matter in sets of total charge zero, simply because all fields of that system are neutral under the graviphoton charge. In particular, the minimum number of RN black holes that can be produced is two. A pair of oppositely charged black holes now ``oxidizes" to a 5D string joining the two charges, which extends inside $AdS_5$ only for a finite amount extent. In particular, the $z$ coordinate remains finite along the string and thus it never comes close to the Poincar\'e horizon, where the metric describing a constant-tension string is necessarily 
singular~\cite{clp,fp}. 

Section 3 is devoted to solving a different puzzle. It arises in theories that contain an infinite number of high spin massless fields in $AdS_5$ These theories where proposed long ago by Fradkin and Vasiliev~\cite{fv} (see~\cite{review} for a recent review). If these theories admit a 4D brane with critical tension, then $AdS_5$ could be capped at $z=\epsilon$ and the induced metric on the 4D brane would be flat. As we shall show, in this cutoff $AdS_5$ space, the radial wave function of high spin fields is normalizable at $k^2=0$; then, the same construction giving rise to 4D dynamical gravity in RS2 would give rise to {\em massless} dynamical high spin fields in Minkowski 4D space. It is easy to show that these fields interact with gravity through their stress energy tensor, i.e. according to the universal interaction implied by the principle of equivalence; but this is in contradiction with several well known no go theorems, in particular with~\cite{p}! The solution we propose is a rather straightforward one. Namely, exactly as predicted by the no go theorem of~\cite{p}, the spurious (gauge) modes of all states of spin higher than two do not cancel
from the cubic vertex describing their interaction with the graviton. So, the gauge symmetry ensuring decoupling of the unphysical modes of spin $s>2$ states is anomalous. To make the boundary 4D theory consistent, one must add a mass term for all such fields. From the point of view of the dual boundary $O(N)$ theory~\cite{kp} this corresponds to the fact that the currents of the free $O(N)$ model are no longer conserved in the presence of dynamical gravity and so acquire anomalous dimensions. In a Randall-Sundrum setup, a 4D mass term is an allowed, local boundary counterterm  of the 5D action. 

\section{N=4 Randall Sundrum Supergravity}
\subsection{Perturbative Neutral States}
N=8 5D gauged supergravity~\cite{grw,ppvn} has 32 supercharges that decompose into the 16 Poincar\'e and 16 conformal supercharges of 4D N=4 supersymmetry. The N=8 graviton multiplet, upon dimensional reduction to 4D, decomposes into N=4 supermultiplets as follows~\footnote{Multiplicity and spin of supermultiplet fields  are labeled as $m (s)$.}:  one graviton multiplet $[1(2), 4 (3/2), 6 (1), 4 (1/2), 2 (0)]$, four gravitino multiplets $[1 (3/2), 4 (1), 7 (1/2),  8 (0)]$,
and six vector multiplets $[1 (1), 4 (1/2), 6 (0)]$. 

On $AdS_5$, gauged N=8 supergravity~\cite{grw,ppvn}, is holographically dual to N=4 SYM. When $AdS_5$ is regulated in the UV by
adding a boundary near the brane, some of the bulk 5D modes give rise to dynamical 4D fields and some do not. We want to understand which one of the potential zero modes becomes dynamical and which one does not.

The 5D Anti de Sitter metric in Poincar\'e coordinates, given above in eq.~(\ref{m1}); the brane is at $z=\epsilon \ll L$. 

On the brane we expect only the graviton supermultiplet of N=4 Poincar\'e supergravity to propagate, because it is dual to the supercurrent multiplet of N=4 SYM. Yet a naive dimensional reduction yields many more fields. 

Some of the fields that would be present in naive dimensional reduction are pure gauge on the cutoff $AdS_5$ space.
These include the $A_5$ component of the 5D graviphoton (i.e. the $SU(4)$ gauge fields) as well as the $g_{55}$ and $g_{\mu 5}$ components of the
metric. Yet, the 42 scalars of the 5D N=8 graviton multiplet can neither be gauged away nor eliminated by boundary
conditions, since the latter are Neumann at $z=\epsilon$ in RS2. 

Rather, these scalars are nondynamical because their kinetic term diverges. One can be more general and examine
under which conditions a 5D $p$ form gives a 4D normalizable kinetic term on UV cutoff $AdS_5$. 

We are interested in 4D massless modes arising out of a $p$ form $A$. Its 5D profile is determined by solving the equation
\beq
d*dA + m^2 *A=0, \qquad A \mbox{ independent of } x^\mu .
\eeq{m2}
The solution is a pure power in $z$. The corresponding massless mode is 
$A(x,z)=(z/L)^E a_{\mu_1,...,\mu_p}(x)dx^{\mu_1}\wedge .... dx^{\mu_p}$. The power $E$ is determined by the mass $m$. 
It is more convenient to express it in terms of the conformal scaling dimension of the 4D current dual to $A$, called $\Delta$. The relation is $E=4-\Delta -p$~\footnote{Under $x^\mu \rightarrow \lambda x^\mu$, $z\rightarrow \lambda z$ the source $a(x)$ scales as 
$a(x)\rightarrow \lambda^{E+p}a(\lambda x)$, so $E+p$ equals the dimension of $a(x)$, that is 4 minus the dimension of the dual current.}.

The 4D kinetic term of the field $a(x)$ follows from the 5D kinetic term:
\beq
\int dA \wedge * dA = \int_\epsilon^\infty dz \left({L\over z}\right)^5 \left({z\over L}\right)^{2E+2p+2} \int_4 d_4 a \wedge _4 *_4 d_4 a + .... \, .
\eeq{m3}
The ellipsis denote contributions other than the 4D kinetic term; the subscript $_4$ denotes 4D quantities (integrals, derivatives, Hodge duals etc.).

A dynamical field requires a finite kinetic term; by recalling that $E=4-\Delta -p$ we see that this means $\Delta > 3$ for all
$p$. 

The 42 N=8 scalars decompose into $SU(4)$ representations as $10_C + 20_R  +1_C$~\cite{ffz}. Those belonging to the real 
representation $20_R$ have $\Delta=2$, those belonging to the complex $10_C$ have $\Delta=3$, while the complex singlet $1_C$ has $\Delta=4$. Our argument shows that only the latter becomes dynamical in N=4 supersymmetric
RS2. Because we know that N=4 supersymmetric boundary conditions exist for RS2, all fields in multiplets
containing the non-dynamical scalars in the $10_C$ and $20_R$ must disappear. 

It is easy to find which multiplet disappears. It cannot be the graviton multiplet, and indeed the complex scalar of that multiplet is the dynamical $\Delta=4$ $SU(4)$ singlet. We are left with the gravitino and vector multiplets. 

The 32 scalars of the 4 gravitino multiplets decompose into two $6_R$ and one $10_C$ of $SU(4)$; the 36 scalars of 
the 6 vector multiplets decompose into one real singlet, a $15_R$ and a $20_R$. 

So, the $\Delta=3$ scalars belong to the gravitino multiplet while the $\Delta=2$ scalars belong to the vector multiplet.

As we said earlier, all the states in N=4 multiplets containing nondynamical scalars must be nondynamical. We can check this for instance by noticing that the 5D $SU(4)$ gauge vectors have $\Delta=3$ so their kinetic term diverges logarithmically in $z$~\cite{g} at large $z$.~\footnote{This logarithmic divergence has a nice interpretation in the dual N=4 SYM; namely: the gauged $SU(4)$ R symmetry is IR free.~\cite{ahpr}}

Among other scalars that would be present in the naive dimensional reduction, those coming from the 5D antisymmetric 2-forms are worth special attention~\cite{dl}. In gauged supergravity, antisymmetric 2-forms are in a $(2,6)$ of $SL(2)\times SU(4)$. The $SL(2)$ interchanging them is the same arising from the axion-dilaton duality of 10D.  The antisymmetric 2-forms are massive, so they do not give rise to 4D scalars. They do not give rise to massive 4D vectors either, unlike in flat
space compactifications. 

To see why, let us call $B^{iA}$ the 2-forms, with $A$ the index of the $6$ of $SU(4)$ and $i$ the index of the $2$ of
$SL(2)$. The 2-form 5D action is first order, so instead of eq.~(\ref{m3}) we have:
\beq
\int B^{iA} \wedge d B^{jA} \epsilon_{ij} + g B^{iA}\wedge * B^{iA}  ,
\eeq{m4}
where $g$ is the coupling constant of the $SU(4)$ gauge group.
The ``mass" term breaks $SL(2)$ and vanishes in the limit that the $SU(4)$ gauge coupling constant goes to zero (and therefore $L\rightarrow \infty$). 

One of the two forms (say $B^{2A}$) can be considered as a non-dynamical multiplier and eliminated through its equations of motion.
The resulting action for the other form is ($B^{1A}\equiv B^A$)
\beq
\int {1\over g} dB^A \wedge * dB^A + g B^A \wedge * B^A .
\eeq{m5}
We know that in N=8 gauged supergravity the conformal dimension associated to $B^A$ is $\Delta=3$, so the argument given below eq.~(\ref{m3}) says that the kinetic term of its zero mode diverges. This is good because a divergent kinetic
term implies $B^A=dC^A$, where $C^A$ is a one form. It possesses a zero mode $C^A= (z/L)^E c^A(x)_\mu dx^\mu$. The coefficient $E$ here is the same as
in $B$, i.e. $E= -1$. The induced 4D action describes 6 propagating {\em massless} vectors:
\beq
\int_\epsilon^\infty dz \left({L\over z}\right)^5 \left({z\over L}\right)^{2E+4} {1\over g} \int_4 d_4 c^A \wedge_4 *_4 d_4c^A = {L^3\over 2g\epsilon^2}
\int_4 d_4 c^A \wedge_4 *_4 d_4c^A.
\eeq{m6}

These are precisely the 4D N=4 graviphotons.

Fermion zero modes too can become dynamical only when their 4D kinetic term is finite. Since Anti de Sitter space in Poincar\'e coordinates is conformal to flat space, the $AdS_5$  5-beins are
\beq
e^5={L\over z} dz, \qquad e^a= {L\over z} \delta^a_\mu dx^\mu, \qquad a=0,1,2,3.
\eeq{m6a}

A Fermionic zero mode is $\psi(z,x)= z^E \phi(x)$, where $\phi(x)$ is a 4D Fermion.
In this case too $E=4-\Delta -p$ ($p=0$ for a spin 1/2 Fermion and $p=1$ for the gravitino). 

The 5D kinetic term reads
\bea
\int d^5x e \bar{\psi} \Gamma^a e_a^M D_M  \psi && \mbox{for spin 1/2},\nonumber \\ \int d^5x e \bar{\psi}_a \Gamma^{abc} e_b^M D_M \psi_c  && \mbox{for the gravitino}.
\eea{m7}
Since the $\Gamma^a$ are flat space 5D gamma matrices the 4D kinetic term of the zero modes is in both cases proportional
to $\int_\epsilon^\infty  dz (z/L)^{4-2\Delta}$ and converges only when $\Delta>5/2$. 

The conformal weight of all Fermion zero modes of N=8 5D gauged supergravity is known. The gravitino and a spin 1/2
Fermion, both in the $4_C$ of $SU(4)$ have $\Delta=7/2$, while a spin 1/2 Fermion in the $20_C$ has 
$\Delta=5/2$~\cite{ffz}.
So, no component of the latter becomes dynamical in RS2 while half of the former do. The other half is removed by the boundary conditions at $z=0$. For both spin 1/2 Fermions and gravitini it  has the schematic form $\gamma^5 \psi^I=\alpha\psi^I$, $I=1,2,3,4$, $\alpha^2=1$ (see~\cite{dl} for more details in the N=2 case).

\subsection{Extremal Charged Black Holes}
Extremal charged black holes are 4D solutions to Einstein's equations that can be uplifted to 5D through 
``oxidation"~\cite{clp}. Their key property is that they are charged under the graviphotons of N=4 4D supergravity. Since the graviphotons are zero modes of 5D antysymmetric tensors, the five dimensional uplift of the black holes is a string. For an isolated black hole, this string must be oriented along the radial Anti de Sitter coordinate $z$. $1/4$ BPS black holes possess an electric charge $p$ under one of the six $U(1)$'s gauged by graviphotons and a magnetic charge $q$ under another one of the $U(1)$'s. In terms of $p,q$ and the asymptotic value of the dilaton $\phi_\infty$, the  resulting 5D metric and dilaton are~\cite{clp}
\bea
ds^2&=&{L^2 \over z^2} \left[ dz^2 -(H_1 H_2)^{-1} dt^2 +H_1 H_2 (dr^2 +r^2 d\Omega_2^2)\right], \nonumber \\
e^{-2\phi}&=&{H_1\over H_2}, \qquad H_1=e^{-\phi_\infty} + {p\over r}, \qquad H_2=e^{\phi_\infty} + {q\over r}.
\eea{m8}

As pointed out already in~\cite{clp}, this metric has a naked singularity at the Poincar\'e horizon $z=\infty$. Indeed, any metric of the form $ds^2= L^2 z^{-2} d\tilde{s}^2$ is singular at $z=\infty$ whenever the Weyl tensor of the metric  $d\tilde{s}^2$, $\tilde{W}_{\mu\nu\rho\sigma}$, is not identically zero~\cite{clp}. Naked singularities are not necessarily fatal in the context of Randall-Sundrum models, because it may be possible to interpret them in terms of physical properties of the dual CFT plus gravity. Examples of acceptable naked singularities were given in~\cite{gpz,gppz}. They were studied in depth in~\cite{gu}. So, even for extremal black holes naked singularities could be acceptable, or other $5D$ solutions could exist, besides that coming from ``oxidation." All this was briefly discussed at the end of~\cite{fp}. 

Here, we would like to present another, more obvious solution to the singularity puzzle.

Isolated, charged black holes are inherently singular, as all isolated objects charged under a massless $U(1)$ are. In a quantum field theory, they live in a superselection sector  different from that of zero-charge states. This is a bit formal but it reflects the simple fact that neutral matter can only produce multiple charged objects, with vanishing total charge. In
particular, since neither the N=4 CFT nor the N=4 supergravity multiplet states carry any graviphoton charge, an extremal black hole cannot be produced in isolation. The smallest number of black holes one can produce in collisions of neutral matter is of course two, with opposite charges. But this configuration, i.e. a black hole with charge vector $\bf{q}$~\footnote{In the charge vector we collect all 12 electric-magnetic graviphoton charges.}
separated by a finite distance $D$ from another black hole with charge vector $-\bf{q}$, has a different 5D uplift than the
isolated black hole metric~(\ref{m8}).  The exact uplift of such {\em non static} solution is unknown; indeed, even the 4D solution can only be studed numerically, since it is time dependent and non supersymmetric. Yet the qualitative feature of the 5D uplift is easy to find. First of all, there must be a sting-like configuration joining the two charges. Furthermore, when the distance between the two opposite-charge black holes is much larger than their horizon radius $D\gg R_S$, away from a cylindrical region of radius $R_S$ surrounding the string both the 4D metric and its 5D uplift are close to flat 4D Minkowski and $AdS_5$, respectively. So, to understand the qualitative behavior of the solution it is legitimate to study the dynamics of a 5D string terminating on the 4D charges in the probe limit, which neglects the string backreaction on the metric. 
In this limit the strings settle to a minimum energy configuration made of a {\em single} string joining the two charges. The string probes the $AdS_5$ bulk, but only up to a finite distance $z\propto D$. 

Explicitly~\cite{m}, one can set one of the charges at $x^1=x^2=x^3=0$ and the other at $x^1=D$, $x^2=x^3=0$. 
If one parametrizes the string as $z=z(x,t)$ (here and afterwards $x\equiv x^1$), then the induced metric on the string world sheet is $ ds^2_{ws}=(L/z)^2[(1+x'^2)dz^2 -(1-\dot{x}^2)dt^2]$. The prime denotes a derivative w.r.t. $z$ and the dot a derivative w.r.t. the Poincar\'e patch time $t$. The induced action for the string is 
\beq
S=T\int dz dt L^2 z^{-2} \sqrt{ (1+x'^2)(1-\dot{x}^2)}.
\eeq{m9}
The string tension $T$ is constant on the Anti de Sitter background; it can be read off the kinetic term of the antisymmetric tensors $B^{iA}$ (see e.g.~\cite{gppz2}).
When we keep the black hole position fixed in 4D, the equations of motion following from action~(\ref{m9}) admit a static
solution. It obeys $[(L/z)^2x'(1+x'^2)^{-2}]'=0$. This equation can be integrated to $x'=\pm (Ez/L)^2[1-(Ez/L)^4]^{-1/2}$ with $E$ a positive constant. This solution extends into $AdS_5$ only up to $z_{max}=L/E$. The constant $E$ is fixed by the condition
\beq
D=2\int_0^{L/E} dz (Ez/L)^2[1-(Ez/L)^4]^{-1/2}.
\eeq{m10}
This equation implies $z_{max}\propto D$.

So, while the 5D uplift of a single black hole is a string extending along $z$ up to the problematic Poincar\'e patch horizon, the uplift of a physical configuration of black holes, with zero total charge, uplifts to a string which only extends up to a finite $z_{max}$. 
This string is localized in all five dimensions, so even when backreaction is taken into account, it can only generate a metric than decays rapidly towards pure $AdS_5$ at large $z$, withtout any naked singularity at $z=\infty$.
A singularity arises only in the limit that $D\rightarrow \infty$, when 4D black holes become infinitely separated and 
$z_{max}\rightarrow \infty$. This singularity then reflects the physical fact that charged states at infinite separation are
singular also in the 4D field theory. For instance, they give infinite cross sections because of the classical Rutherford scattering divergence. 

We expect more complex configurations of three or more black holes also to uplift to 5D multi-pronged strings configurations, which extend in the $z$ coordinate only up to a maximum finite value and therefore generate metrics regular at the Poincar\'e patch horizon.

\section{RS2 Meets Vasiliev}
In Anti de Sitter spaces it is possible to construct interacting theories of massless high spin states. These theories were pioneered by Fradkin and Vasiliev in~\cite{fv}. A modern review on this vast subject is~\cite{review}. Some of these theories were conjectured to be dual to $O(N)$ models in~\cite{kp}. In particular, an $AdS_4$ theory containing massless particles of  arbitrary even spin --$hs(4)$ in the classification of ref.~\cite{ess}-- was conjectured to be dual to $N$ free scalars in 3D. 
A similar duality is believed to hold between $AdS_5$ Vasiliev theories and $N$ free massless scalars in 4D ($\phi^I$).
Under this duality, the boundary value of a massless spin-$s$ $AdS_5$ field sources the spin-$s$ conserved current 
\beq
J_{\mu_1 .... \mu_s}= \sum_{I=1}^N \phi^I \stackrel{\leftrightarrow}{\partial}_{(\mu_1}.....\stackrel{\leftrightarrow}{\partial}_{\mu_s)_T}\phi^I , \qquad s>2, \qquad J_{\mu\nu}=T_{\mu\nu},
\eeq{m11}
where $(\mu_1... \mu_s)_T$ denotes the symmetric-traceless tensor product and $T_{\mu\nu}$ is the traceless stress energy tensor.

A very nontrivial check of this conjecture was performed in~\cite{gy}, where current correlators in the free 4D CFT were reproduced by computing Witten diagrams~\cite{w} in the dual Vasiliev theory~\footnote{Ref.~\cite{gy} also computes current correlators in critical $O(N)$ model, finding again agreement with bulk calculations.}. 

A puzzle arises now if we cut off $AdS_5$ space in a Vasiliev theory; in other words, if we extend the  RS2
construction to high spin theories. 

Even for Vasiliev theories, the crucial  property of the Randall-Sundrum construction holds; namely, that all states with $k^2=0$ and a normalizable 5D wave function appear as massless particles in 4D Minkowski space. 
These particles interact with each others and with the graviton, simply because the n-point correlators of the currents~(\ref{m11}) do not vanish. Moreover, they are invariant under all the spin-$s$ gauge symmetries needed to decouple spurious polarizations, because this is the defining property of the Fradkin-Vasiliev construction~\cite{fv}. Finally, the interaction with the graviton obeys the principle of equivalence; this follows from the fact that correlators of high spin currents with $T_{\mu\nu}$ obey standard Ward identities. {\em but this contradicts old~\cite{wein,des} and new~\cite{p} no go theorems!}.  

Something in the previous line of reasoning must be incorrect. One possibility is that high spin fields do not give normalizable modes at $k^2=0$. This possibility though is easily excluded using the arguments used for antisymmetric forms in the previous section.  Indeed, even for a spin-$s$ field, $\psi_{m_1,..m_s}(z,x)$, $m_i=0,1,2,3,4$, it is still true that the $k^2=0$ wave function must behave as $\psi_{m_1,..m_s}(z,x)=z^E \hat{\psi}_{m_1,..m_s}(x)$; moreover, it is still true that their $m_i=4$ components can be set to zero with a gauge choice. Finally, since $\psi_{\mu_1,..\mu_s}(x)$ is a rank-$s$ covariant tensor, it is also still true that the relation between $E$ and the scaling dimension $\Delta$ is 
$E=4-\Delta -s$. Thus, the 5D kinetic term of the massless mode for rank-$s$ field becomes
\bea
\int d^4x dz \sqrt{-g} g^{m_1 n_1}.... g^{m_{s+1} n_{s+1}} \partial_{m_1}\psi_{m_2... m_{s+1}} (z,x)\partial_{n_1} \psi_{m_2... m_{s+1}}(z,x) +.... & & \nonumber \\
\propto  \int_\epsilon^\infty dz z^{2s-3}z^{8-2s-2\Delta}\int d^4x\partial_{\mu_1}\hat{\psi}_{\mu_2... \mu_{s+1}} (x)\partial^{\mu_1}\hat{\psi}^{\mu_2... \mu_{s+1}} (x) +.....&&
\eea{m12}
The dimension $\Delta$ is $s+2$, as it is evident from duality of $\psi_{m_1,..m_s}$ with the currents~(\ref{m11}) and from the fact that the scaling dimension of a 4D scalar is one; thus, the integral in $dz$ in eq.~(\ref{m12}) converges for all spins $s>1$. 

The problem arises with interactions, as it may have been expected; specifically with the spin-$s$ gauge invariance of interactions, starting at the first nontrivial order (cubic in the fields). When we stated that the Fradkin-Vasiliev construction ensured gauge invariance of the full interacting action under spin-$s$ gauge transformations, we implicitly assumed that the gauge 
transformations were normalizable near $z=0$. This is the correct boundary condition for $AdS_5$, but not for $AdS_5$
cut off at $z=\epsilon$. In the latter case, it is precisely the non-normalizable gauge transformations that become the gauge
transformations of 4D massless particles. It is they that ensure that 4D spurious polarizations decouple. It is also they that may not leave the action invariant, because for them it is no longer legitimate to integrate by part and disregard boundary terms. 

To be specific, let us start by writing down the inhomogeneous part of the spin-$s$ gauge transformation
\beq
\delta\psi_{m_1,..m_s}(z,x)=D_{(m_1}\epsilon_{m_2... m_s)_T} (z,x).
\eeq{m13}
The 4D gauge transformations are those that leave the field $\psi_{m_1,..m_s}(z,x)=z^E \hat{\psi}_{m_1,..m_s}(x)$ in the
gauge $\psi_{4, m_2..m_s}=0$. This condition constrains the gauge parameter to have the form
\beq
\epsilon_{(m_1... m_{s-1})_T}=z^{2-2s}\hat{\epsilon}_{(m_1... m_{s-1})_T}. 
\eeq{m14}

Next, decompose the bulk action into a free quadratic part plus an interacting part. The quadratic part is schematically
\beq
S_2=\int d^4x dz \psi(z,x)(L \psi)(z,x) + \int d^4x \hat{\psi}(x) (B\hat{\psi})(x).
\eeq{m15}
The kinetic term of the bulk quadratic action has been denoted here by $L$, while all boundary terms needed to enforce Neumann boundary conditions have been called $B$. These terms ensure that when $\hat{\psi}$ obeys the 4D equations of motion of a free massless spin-s particle, then the action is stationary {\em even under variation that do not vanish at the boundary} $z=\epsilon$, such as those given by eq.~(\ref{m13}) with gauge parameter~(\ref{m14}).

Interactions arise first at cubic order. One universal interaction term that is always present involves two spin-$s$ fields and a metric fluctuation $h_{mn}(z,x)\equiv L^{-2} z^{2} g_{mn} -\eta_{mn}$. Schematically the action is
\beq
S=S_2+S_3, \qquad S_3 = \int d^4x dz V[\psi(x),\psi(x),h(x)].
\eeq{m16}
The local cubic interaction $V[\psi(x),\psi(x),h(x)]$ is a sum of two terms: $V_m+V_{FV}$. The first one comes from covariantizing the 
action~(\ref{m16}) using the minimal coupling procedure and expanding to linear order in $h_{mn}$. The second is the Fradkin-Vasiliev (FV) vertex~\cite{fv}, which is needed to ensure consistency of the equations of motion to cubic order (see~\cite{bls} for a recent derivation of this vertex). 

Now a crucial observation is that if the fields $\psi$ are on shell, i.e. if they obey both the 5D bulk equations and the 4D equations of motion, then the quadratic action is invariant under arbitrary variations $\delta \psi$. 
So, under a full non-linear gauge variation and up to quadratic  order in the fields, the change in $S_2+S_3$ reduces to the change in $S_3$:
\beq
\delta S = \int d^4x dz \left\{V[\delta\psi(x),\psi(x),h(x)] + V[\psi(x),\delta \psi(x),h(x)]\right\},
\eeq{m17}
with $\delta\psi$ given in eq.~(\ref{m13}).

To find out where the problem lays with the would be massless high spin states in the Randall-Sundrum construction applied to Vasiliev's theories, we consider now the explicit case of spin 3 particles. 

The cubic FV vertex for two $s=3$ particles and a graviton can be found e.g. 
in~\cite{bls} eq. (14). It simplifies dramatically in the gauge $\psi_{4 mn}(z,x)=0$, $h_{4m}(z,x)=0$, especially when the $s=3$ field $\psi_{\mu\nu\rho}$ is on shell and  the metric fluctuation $h_{\mu\nu}$ is independent of $z$. These are the field configurations we need to show the problem with interactions. On such configuration  the FV vertex contains only one term involving derivatives w.r.t. $z$ [compare with~\cite{bls} eq. (14)]:
\beq
\int d^4x dzV_{FV}[\psi,\psi,h]= -{3\over \Lambda} \int d^5x dz {L^5\over z^5} w_{\alpha\beta\gamma\delta}D^z \psi^{\mu\alpha\beta}D_z\psi^{\gamma\delta}_\mu. 
\eeq{m18}
Here $\Lambda$ and $w^{\alpha\beta\gamma\delta}$ are, respectively,  the 5D cosmological constant and the linearized Weyl tensor.
Inserting the gauge variation~(\ref{m13}) into eq.~(\ref{m18}), the only way to cancel the resulting
 term is to integrate by part in $dz$. Integration
by part produces a term that cancels against lower-dimension terms using the free $\psi$ equations of motion. But it also produces the  boundary term
\beq
-{3\over \Lambda}\int d^4x {L^5\over \epsilon^5} w_{\alpha\beta\gamma\delta}D^{(\mu}\epsilon^{\alpha\beta)_T}
\stackrel{\leftrightarrow}{D}_z\psi^{\gamma\delta}_\mu .
\eeq{m19}
This term would have been zero on a metrically complete $AdS_5$ space and for a normalizable gauge variation. On cutoff $AdS_5$ instead, it does not vanish. Moreover, since the only field that is not on shell with respect to the 4D equations of motion is the metric fluctuation, the only chance to cancel~(\ref{m19}) is
by a {\em local} variation of $h_{\mu\nu}$. For this to be possible, eq.~(\ref{m19}) would have to  vanish when
$h_{\mu\nu}$ is on shell. But eq.~(\ref{m19}) is proportional to the Weyl tensor, which does not vanish on shell!

{\em Therefore, the gauge symmetry~(\ref{m13}) is anomalous to first order in the gravitational interactions}.

This is in exact agreement with the findings of ref.~\cite{p}. As for any anomalous gauge symmetry, the only way to
escape algebraic inconsistency is for the high spin field to acquire a mass. From the point of view of the dual boundary $O(N)$ theory, the need for a mass term follows from the fact that the currents of the free $O(N)$ model are no longer conserved in the presence of dynamical high spin fields. Indeed, just the presence of a dynamical graviton implies that the  high spin currents are only covariantly conserved and so can acquire anomalous dimensions.
Luckily, a mass counterterm is natural in the RS2 construction, since it can be introduced simply by modifying the term 
$B$ in eq.~(\ref{m15}). So, a RS2-type construction is possible even for Vasiliev high spin theories but the resulting  dynamical gravity plus matter in 4D Minkowski space  cannot contain massless particles of spin higher than 2.

\subsection*{Aknowledgements}
This material is based upon work supported in part by the National Science Foundation under Grant No. 1066293 and the hospitality of the Aspen Center for Physics.  

S.F. research is supported by ERC Advanced Investigator Grant n.226455 {\em Supersymmetry, Quantum Gravity and Gauge Fields (Superfields)}.

M.P. would like to thank CERN for its kind hospitality. M.P. is supported in part by NSF grant PHY-0758032, and
by ERC Advanced Investigator Grant n.226455 {\em Supersymmetry,
Quantum Gravity and Gauge Fields (Superfields)}.

\end{document}